\documentclass[twocolumn,showpacs,preprintnumbers,amsmath,amssymb,prl]{revtex4}

\usepackage{graphicx}
\usepackage{dcolumn}
\usepackage{bm}

\begin{document}


\title{Quantum Versus Jahn-Teller Orbital Physics in YVO$_3$ and LaVO$_3$}

\author{Zhong Fang$^{1}$, and Naoto Nagaosa$^{2,3}$}

\affiliation{
$^1$Institute of Physics, Chinese Academy of Science, Beijing 100080, China\\
$^2$Correlated Electron Research Center (CERC), AIST Tsukuba
Central 4, 1-1-1 Higashi, Tsukuba, Ibaraki 305-8562, Japan;\\
$^3$Department of Applied Physics, University of Tokyo, 7-3-1, Hongo,
Bunkyo-ku, Tokyo 113-8656, Japan
}

\date{\today}

\begin{abstract}
We argue that the large Jahn-Teller (JT) distortions in YVO$_3$ and
LaVO$_3$ should suppress the quantum orbital fluctuation.  The unusual
magnetic properties can be well explained based on LDA+$U$
calculations using experimental structures, in terms of the JT
orbital. The observed splitting of the spin-wave dispersions
for YVO$_3$ in C-type antiferromagnetic state is attributed to the 
inequivalent VO$_2$ layers in the crystal structure, 
instead of the ``orbital Peierls state''. 
Alternative stacking of $ab$-plane exchange couplings
produces the $c$-axis spin-wave splitting, thus the spin system is highly three
dimensional rather than quasi-one-dimensional. Similar splitting is also
predicted for LaVO$_3$, although it is weak.
\end{abstract}

\pacs{71.27.+a, 75.10.-b, 71.70.-d, 75.30.-m}

\maketitle

An essential subject, which is responsible for the rich physics in
transition-metal oxides, is the orbital degree of freedom (ODF) and
its interplay with the spin, charge and lattice degrees of
freedom~\cite{Imada,GKR,Khomskii}.  For cubic perovskites, quantum
orbital fluctuation (QOF) would be expected due to the degeneracy in
orbital sector, and a particular spin and orbital ordered phase can be
selected by maximizing the energy gain from the
QOF~\cite{Keimer,Khaliullin,motome,YVO1}. On the other hand,
Jahn-Teller (JT) distortions will lift the orbital degeneracy, and
suppress the QOF~\cite{Mizokawa1,Sawada,Cwik,Hemberger}.  The
competing QOF versus JT physics has thus been the central issue for
the recent discussions on $t_{2g}$ perovskites. A typical example is
LaTiO$_3$, which shows much reduced ordered moment from that expected
by mean field theory, and isotropic spin dynamics~\cite{Keimer}. At
one side, the relevance of QOF was emphasized~\cite{Khaliullin} by
neglecting the crystal field effects and treating the systems
essentially as cubic. On the other hand, this scenario is questioned
by the recent examinations, which demonstrated that those unusual
magnetic properties can be understood from polarized orbital caused by
the lattice distortions~\cite{Cwik,Hemberger,Imada3,Kiyama}, and that
the predicted orbital entropy from QOF was not
observed~\cite{entropy}.

YVO$_3$ and LaVO$_3$ are $t_{2g}$ perovskites with two localized 3$d$
electrons per V. Early experimental results showed complicated phases
for those compounds~\cite{LVO1,YVO2,Ren}. First, LaVO$_3$ has
G-type~\cite{Note2} JT distortion and the C-type antiferromagnetic
(AF) state below 140K~\cite{LVO1}. YVO$_3$ does the same between
77K$<T<$116K, but shows the C-type JT distortion, and the G-type AF
state below 77K~\cite{YVO2}. Very recently, an unusual magnetic
structure and dynamic in YVO$_3$ was reported~\cite{YVO1}, especially
a splitting of the $c$-axis spin-wave dispersions was observed for the
intermediate temperature (77K$<T<$116K) phase of YVO$_3$, where the
$c$-axis lattice dimerization is vanishingly weak~\cite{YVO2,FIR}. The
QOF was thus argued for those compounds by neglecting the strong JT
distortions present (about 2$\sim$4\%). Theoretically, by treating the
system as quasi one-dimensional (1D), the spin-orbital superexchange
model was analyzed, and an ``orbital-Peierls state'' due to the
formation of orbital singlet was proposed~\cite{YVO1}. However, the
spin system can not be regarded as 1D, while the orbital system is due
to the destructive interference of the interchain exchange processes
\cite{motome}.  Furthermore, from this QOF picture it is hard to
understand the observed large JT distortion and its clear temperature
dependence~\cite{YVO2}. It is therefore an interesting and challenging
problem to judge the underlying physics here, especially by
first-principles calculations.

In this letter, we will present firm evidences for the crucial role of
lattice distortion for YVO$_3$ and LaVO$_3$. The experimentally
observed spin orderings (SO) can be systematically explained by our
LDA+$U$ calculations, in terms of the JT orbital. The ``unexpected
features'' of spin wave in YVO$_3$~\cite{YVO1}, namely, 1) splitting
of $c$-axis spin wave; 2) $|J_c|>|J_{ab}|$ ($J_{c}$ and $J_{ab}$ are
$c$-axis and $ab$-plane exchange couplings, respectively), can be
naturally explained from the structural point of view.  Similar
picture is also predicted for LaVO$_3$. We further point out that the
theoretical simplification to treat the system as quasi 1D is lack of
firm bases.

The calculations are done based on the LDA+$U$ scheme~\cite{LDAU} in
plane-wave pseudopotential method~\cite{PP}. For such well defined
insulating systems with long range ordering, LDA+$U$ method typically
can give reasonable results.  The parameter ~\cite{JPC} $U_{\rm
eff}$=3.0eV is used to reproduce the experimental band gaps
properly~\cite{gap}.  For all the discussions, we use the unit cell
with $a\approx b \approx c/\sqrt{2}$, which includes four V sites
(sites 1 and 2 in one layer, with sites 3 and 4 on top of 1 and 2,
respectively), and define the local axes $x, y, z$ as the [110],
[$\bar{1}$10], [001] directions of the unit cell. We performed ground
state calculations for three structures, YVO$_3$ at 65 K, at 100K, and
LaVO$_3$ at 10K~\cite{LVO1,YVO2}. Four magnetic structures (i.e., FM
and A-, C-, G-type AF states~\cite{Note2}) were calculated for each
fixed experimental structures.

\begin{table}
\caption{The calculated total energies $E$(meV/f.u.), magnetic moment
$M$($\mu_{B}$/site), and band gap $E_g$(eV), for various compounds in
different magnetic states. For YVO$_3$ (100K) and LaVO$_3$ (11K), two
magnetic moments are given for two inequivalent layers. Bold numbers
correspond to most stable magnetic state.}
\begin{ruledtabular}
\begin{tabular}{c|c|c|c|c|c}
     & &FM &A-AF &C-AF &G-AF   \\ \hline
     &$E$ &45.9 &27.2 &12.8 &{\bf 0.0} \\ \cline{2-6}
YVO$_3$ &$M$ &1.75 &1.72 &1.70 &{\bf 1.68}  \\ \cline{2-6}
(65 K)  &$E_g$ &0.8  &1.0  &0.8  &{\bf 1.2} \\ \hline \hline
     &$E$ &16.7 &27.6 &{\bf 0.0} &19.1 \\ \cline{2-6}
YVO$_3$ &$M$ &1.75(1.77) &1.72(1.75) &{\bf 1.70(1.72)} &1.68(1.70)  \\ \cline{2-6}
(100 K) &$E_g$ &0.8  &1.1  &{\bf 1.2}  &1.0 \\ \hline \hline
     &$E$ &38.3 &42.0 &{\bf 0.0} &23.5  \\ \cline{2-6}
LaVO$_3$ &$M$ &1.74(1.75) &1.72(1.72) &{\bf 1.70(1.71)} &1.69(1.70) \\ \cline{2-6}
(10 K)   &$E_g$ &0.7  &1.0  &{\bf 1.2}  &0.9
\end{tabular}
\end{ruledtabular}
\end{table}

Let us start from the low temperature phase of YVO$_3$, which has
C-type JT distortion~\cite{YVO2}. As shown in Table I, among the four
magnetic structures, the G-AF state is the most stable state,
consistent with experimental observation and earlier
calculations~\cite{Sawada}. The obtained magnetic moment
(1.68$\mu_{\rm B}$) and band gap (1.2eV) are also in excellent
agreement with experimental ones (1.72$\mu_{\rm B}$~\cite{YVO1} and
1.2eV~\cite{gap}), demonstrating the validity of our approaches. The
stabilization of G-AF state can be naturally explained in terms of the
JT orbital as follows. By calculating the occupation numbers, it is
easy to find that the C-type JT distortion stabilize the C-type
orbital ordering (OO), where $yz, zx, yz, zx$ orbital is occupied for
four V sites (1,2,3,4) respectively, and $xy$ orbital is occupied for
all V sites. In this OO pattern, the super-exchange (SE) along the
$c$-axis is expected to be AF due to the ferro-orbital chain
arrangement of $yz$ and $zx$, according to the Goodenough-Kanamori
rules (GKR). There are two contributions, however, for the SE in
$ab$-plane, the FM one due to the nearly orthogonal $yz, zx$ orbital
arrangement, and the AF one coming from $xy$ orbital. The net coupling
will depend on the relative polarization of those orbital. The
calculated orbital occupation numbers ($n_{xy}, n_{yz}$ and $n_{zx}$)
for one of the V sites (shown in Fig.1(a)) clearly show that the
polarization between the $yz$ and the $zx$ states is not so strong,
and the net magnetic interaction is dominated by the AF SE from $xy$
state. As the results, we would expect the AF exchange coupling both
along $c$-axis and in the $ab$-plane. This will explain the ground
state G-type SO.

\begin{figure}
\includegraphics[scale=0.45]{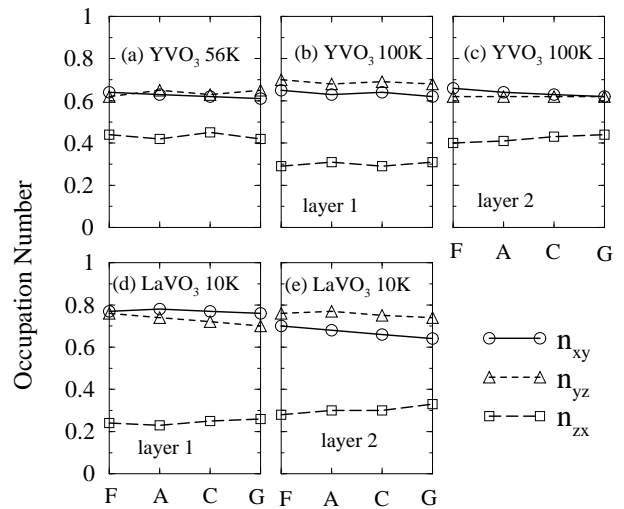}
\caption{The calculated $t_{2g}$ orbital occupation numbers for
different compounds as function of different magnetic orderings. Since
the OO patterns are fixed by the structures, only the occupations for
one of the transition-metal site are shown.}
\end{figure}

Importantly, the calculated OO patterns and magnetic moments are not
sensitive to the SO. For the fixed structure of YVO$_3$ at 65 K, we
always get the C-type OO for the four magnetic states. Furthermore,
even the occupation numbers of each orbital do not change so much for
different magnetic states, as shown in Fig.1. The situation is quite
general for all the examples considered here, and is also true for
LaVO$_3$. This clearly suggests the crucial role of lattice
distortions for those compounds. The experimentally observed JT
distortions are essentially important to reproduce the correct
magnetic orderings.  Giving the facts that all those calculated
magnetic states have basically the same OO pattern, and that the
charge gap $E_g$ (see Table I) is much larger than the spin excitation
energies (typically the order of meV), we can approximately decouple
the spin degree of freedom, and treat it in terms of the Heisenberg
model. Then the exchange interactions can be estimated by mapping the
calculated total energies for each magnetic state, $E$(F), $E$(A),
$E$(C) and $E$(G), to the Heisenberg model. The nearest neighboring
exchange coupling constants are then given by:
\begin{eqnarray}
J_c&=& (1/4S^2) [E({\rm F})-E({\rm G})-E({\rm A})+E({\rm C})]
\nonumber \\ J_{ab}&=& (1/8S^2)[E({\rm F})-E({\rm G})+E({\rm
A})-E({\rm C})]
\end{eqnarray}
where $S$=1 is the moment. For YVO$_3$ at 65K, we got $J_c$=7.8meV and
$J_{ab}$=7.5meV, which is quite isotropic as suggested by
experments~\cite{YVO1}, and can be reasonably compared with the
experimental values as shown in Table II. Up to this stage, we show
that the low temperature phase of YVO$_3$ can be well explained by the
JT orbital physics.

\begin{table}
\caption{The comparison of calculated and experimental spin coupling
constants $J_{c}$ and $J_{ab}$ (meV) for various compounds. The
numbers with parentheses are two inequivalent values as discussed in
the text. The symbols $\dagger$ and $\ddagger$ point to
Ref.~\cite{YVO1} and ~\cite{LVO2} respectively.}
\begin{ruledtabular}
\begin{tabular}{c|c|c|c|c}
     & &YVO$_3$ &YVO$_3$ &LaVO$_3$  \\ 
     & &(65 K)  &(100 K) &(10 K)    \\ \hline
Cal. &$J_c$    &7.8 &-7.2       &-6.5    \\ \cline{2-5}
     &$J_{ab}$ &7.5 &0.8(5.3)   &5.8(7.7)  \\ \hline \hline
Exp. &$J_c$    &5.7$^{\dagger}$ &-2.0(-4.2)$^{\dagger}$ &-4.0$^{\ddagger}$ \\ \cline{2-5}
     &$J_{ab}$ &5.7$^{\dagger}$ &2.6$^{\dagger}$        &6.5$^{\ddagger}$ \\
\end{tabular}
\end{ruledtabular}
\end{table}

For YVO$_3$ at 100K, the experimental G-type JT distortion will
stabilize the G-type OO, in which the $yz$ and $zx$ orbital are
occupied alternatively (antiferro-orbital) along the $c$-axis, instead
of the ferro-orbital chain below 77K. For the four magnetic states, we
all obtained the G-type OO pattern. In such G-type OO state, the
antiferro-orbital chain along $c$-axis will favor FM coupling along
the chain, while the exchange coupling in $ab$-plane are basically the
same as that of YVO$_3$ at 65 K, resulting in the experimental C-AF
state (see Table I)~\cite{Sawada-comment}. Now the question is how to
understand the ``unexpected features'' of spin wave, which were argued
to be the result of orbital dimer formation~\cite{YVO1}. 1)The
$c$-axis spin wave splits into two branches and open a gap;
2)$|J_c|>|J_{ab}|$, while according to GKR, FM SE ($J_c$ in this case)
is generally weaker than AF SE ($J_{ab}$). X-ray diffraction
results~\cite{YVO2} suggested that YVO$_3$ in the C-AF phase has
$P2_1/a$ symmetry, which has no c-axis dimerization. The recent
far-infrared spectroscopy data~\cite{FIR} suggested the possible
lowering of the symmetry group to $Pb11$ or $P_{\bar 1}$. However,
this lowering will not violate our follwoing discussions due to: 1)
the emergence of new phonons is one or two order weaker in
intensity~\cite{FIR} compared with the main modes; 2) the following
arguements are common for all those possible symmetries. A
characteristic point of the lattice structure of YVO$_3$ in C-AF
state, in sharp contrast with the low temperature phase ($Pbnm$ space
group), is the absence of any symmetry operation to transfer one
VO$_2$ layer to the neighboring layer along $c$-direction, resulting
in two inequivalent VO$_2$ layers, which have different amounts of JT
distortion. As the results, we obtain two $J_{ab}$ (=0.8 and 5.3meV)
for two different layers~\cite{Note1}, which stacking along the
$c$-axis alternatively. This is in qualitative difference with the
experimental analysis, in which they assumed alternative $J_c$ but
same $J_{ab}$ to fit the experimental spin-wave. By using the
calculated exchange parameters (alternative $J_{ab}$), our obtained
spin-wave dispersions (shown in Fig.2(B)) definitely shows a
$c$-axis spin-wave splitting, which is comparable in size with the
experimental one (about 5meV). The overall shape of our obtained
spin-wave is also in good agreement with the experimental one. We
conclude that the observed spin-wave splitting is due to the
inequivalent VO$_2$ layers in this compounds. For such strongly
coupled systems both spin and orbital behaviors are essentially three
dimensional. Especially for the spin degree of freedom, the exchange
coupling in $ab$-plane will dramatically affect the spin wave behavior
along $c$-axis. It is generally not suitable to treat the system as
quasi 1D as assumed in previous studies~\cite{YVO1}. The ``orbital
Peierls states'' will be easily suppressed by the increased JT
distortions.

\begin{figure}
\includegraphics[scale=0.4]{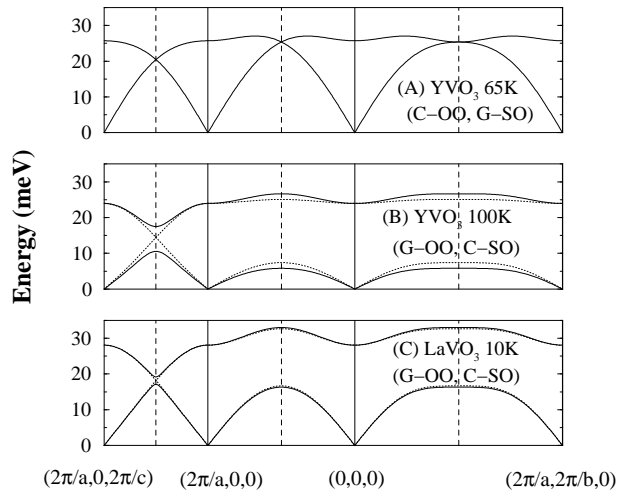}
\caption{The calculated spin-wave dispersions for various compounds in
the Heisenberg model. Solid lines are results using calculated
exchange coupling constants, while dotted lines are that using single
average of two $J_{ab}$.}
\end{figure}

The dramatic difference of two $J_{ab}$ comes from different
distortions of two layers. For YVO$_3$ at 100K, the JT distortion is
about 4\% for one layer (say layer 1), while about 2\% for another
layer (say layer 2). Such a difference will modify the orbital
polarization as shown in Fig.1(b) and (c). The polarization between
the $yz$ and $zx$ orbital for layer 1 is larger than that of layer 2,
while the $n_{xy}$ is almost same for two layers. The larger
polarization tend to enhance the FM coupling coming from the
orthogonal $yz$ and $zx$ occupation in the $ab$-plane, which will
compensate the AF coupling from $xy$ orbital. Therefore, the net AF
coupling are significant suppressed, from 5.3meV for layer 2 to 0.8meV
for layer 1. This fact will also explain why $|J_c|>|J_{ab}|$ in this
case, and again suggest the crucial role of lattice distortions.

Due to the same symmetry, we obtain the similar results for LaVO$_3$
at 10K, i.e., the observed G-type JT distortion will stabilize the
G-type OO robustly, and as the result, the C-AF ground state is
obtained. Our calculated magnetic moment (1.7$\mu_{\rm B}$) is somehow
larger than the early experimental value (about 1.3$\mu_{\rm B}$
~\cite{LVO2}). The $c$-axis spin-wave gap is also predicted in this
case, although it is weak due to the smaller structural difference
between the two VO$_2$ layers (see Table I, II and Fig.1, 2). This
prediction should be verified by future experiments.

Finally, we have two comments for the C-AF phase of YVO$_3$. First,
the experimental moment (about 1.05$\mu_{\rm B}$) is much smaller than
our calculated one (1.70 and 1.72~$\mu_{\rm B}$ for two inequivalent V
sites). This fact has been used as an argument for the QOF
nature. However, from the neutron data~\cite{YVO1}, we find that the
C-type spin diffraction intensity above 77K is far from saturated. By
extrapolating that intensity versus temperature data down to 0K, we
can easily get a increased moment by a factor of 1.7. Therefore, the
reduced magnetic moment could be due to the elevated
temperature. Second, the spin canting of 16 degrees is argued from
experimental side~\cite{YVO1}, while it is not included in our
calculations. Nevertheless, we should say such a canting may
quantitatively affect our results discussed above (such as the
calculated moments for YVO$_3$ and LaVO$_3$ in C-AF state), but not
qualitatively, especially for the main conclusion about the structural
origin of spin wave gap.

In summary, we show that the reported unusual magnetic properties for
YVO$_3$ and LaVO$_3$ can be reasonably explained by a systematic JT
picture from LDA+$U$ calculations. For fixed structures, the obtained
OO patterns are not sensitive to SO, suggesting that any meaningful
orbital fluctuation must be via the phonon degrees of freedom. It will
be an interesting future subject to study the phonon-mediated orbital
fluctuations for those compounds.

\begin{acknowledgments}
The authors acknowledge the valuable discussion with Prof. Y. Tokura,
Prof. K. Terakura, Dr. S. Miyasaka and Dr. Y. Motome. Z. Fang
acknowledges support from NSF of China (No. 90303022 and
10334090).
\end{acknowledgments}


\begin{thebibliography}{}
\bibitem{Imada} M. Imada, {\it et.al}, Rev. Mod. Phys. {\bf
  70}, 1039 (1998).
\bibitem{GKR} J. B. Goodenough, Phys. Rev. {\bf 100}, 564 (1955);
J. Kanamori, J. Phys. Chem. Solids, {\bf 10}, 87 (1959).
\bibitem{Khomskii} K. I. Kugel and D. I. Khomskii,
Sov. Phys. Usp. {\bf 25}, 231 (1982); Sov. Phys. Solid State {\bf 17},
285 (1975).

\bibitem{Keimer} B. Keimer, {\it et al.},
Phys. Rev. Lett. {\bf 85}, 3946 (2000).

\bibitem{Khaliullin} G. Khaliullin, S. Maekawa, Phys. Rev. Lett. {\bf
85}, 3950 (2000); G. Khaliullin,{\it et.al.}, {\it ibid} {\bf 86},
3879 (2001).

\bibitem{motome} Y. Motome, {\it et.al.}, Phys. Rev. Lett. 90, 146602
(2003)


\bibitem{YVO1} C. Ulrich, G. Khaliullin, J. Sirker, and {\it et al.},
Phys. Rev. Lett. {\bf 91}, 257202 (2003); P. Horsch, G. Khaliullin,
and A. M. Ole\'{s}, Phys. Rev. Lett. {\bf 91}, 257203 (2003).

\bibitem{Mizokawa1} T. Mizokawa, A. Fujimori, Phys. Rev. B {\bf 51},
12880 (1995); {\it ibid}, {\bf 54}, 5368 (1996); T. Mizokawa,
D. I. Khomskii, and G. A. Sawatzky, Phys. Rev. B {\bf 60}, 7309
(1999).


\bibitem{Sawada} H. Sawada, N. Hamada, K. Terakura, and T. Asada,
Phys. Rev. B {\bf 53}, 12742 (1996); H. Sawada, K. Terakura,
Phys. Rev. B {\bf 58}, 6831 (1998).
\bibitem{Cwik} M. Cwik, T. Lorenz, J. Baier, and {\it et al.},
Phys. Rev. B {\bf 68}, 60401 (2003).
\bibitem{Hemberger} J. Hemberger, H. A. Krug Von Nidda, V. Fritsch,
and {\it et al.}, Phys. Rev. Lett. {\bf 91}, 66403 2003).

\bibitem{Imada3} M. Mochizuki, M. Imada, Phys. Rev. Lett. {\bf 91},
167203 (2003).
\bibitem{Kiyama} T. Kiyama, M. Itoh, Phys. Rev. Lett. {\bf 91},
167202 (2003).

\bibitem{entropy} V. Fritsch, J. Hemberger, M. V. Eremin, and {\it et
al.}, Phys. Rev. B {\bf 65}, 212405 (2002).

\bibitem{LVO1} P. Bordet, C. Chaillout, M. Marezio, and {\it et al.},
J. Solid State Chem. {\bf 106}, 253 (1993).

\bibitem{YVO2} G. R. Blake, T. T. M. Palstra, and Y. Ren,
Phys. Rev. Lett. {\bf 87}, 245501 (2001); {\it ibid}, Phys. Rev. B
{\it 65}, 174112 (2002); M. Noguchi, A. Nakazawa, S. Oka, and {\it et
al.}, Phys. Rev. B {\bf 62}, 9271 (2000).

\bibitem{Ren} Y. Ren, T. T. M. Palstra, D. I. Khomski, and {\it et
al.} NATURE, {\bf 396}, 441 (1998).

\bibitem{Note2} Following common convention, here we use letters F, A,
C, G to defind four types of ordering. Please refer to~\cite{Imada}
for the detailed definitions.

\bibitem{FIR} A. A. Tsvetkov, F. P. Mena, P. H. M. van Loosdrecht, and
{\it et. al.}, Phys. Rev. B {\bf 69}, 075110 (2004).

\bibitem{LDAU} V. I. Anisimov, J. Zaanen, and O. K. Anderson,
Phys. Rev. B {\bf 44}, 943 (1991); I. V. Solovyev, P. H. Dederichs,
and V. I. Anisimov, Phys. Rev. B {\bf 50}, 16861 (1994).

\bibitem{PP} D. Vanderbilt, Phys. Rev. B {\bf 41}, 7892 (1990).

\bibitem{JPC} Z. Fang, {\it et.al}, J. Phys.: Cond. Matt. {\bf 14},
3001 (2002).

\bibitem{gap} T. Arima, Y. Tokura, J. B. Torrance, Phys. Rev. B {\bf
48}, 17006 (1993); Y. Okimoto, T. Katsufuji, Y. Okada, and {\it et
al.}, Phys. Rev. B{\bf 51}, 9581 (1995); S. Miyasaka, Y. Okimoto, and
Y. Tokura, J. Phys. Soc. Jpn. {\bf 71}, 2086 (2002); Z. Fang,
N. Nagaosa, and K. Terakura, Phys. Rev. B {\bf 67}, 035101 (2003).

\bibitem{Sawada-comment} In the early calculations by Sawada {\it
et. al.}~\cite{Sawada}, the stable C-AF state of YVO$_3$ at
intermediate temperature can not be predicted, in contracdiction with
our present results. This is due to the absense of structural
information presented in Ref.~\cite{YVO2} for Sawada's calculations.

\bibitem{Note1} To separate the contributions from different layers
for the spin exchange interactions, two additional spin
configurations, in which only one of the four spins is flipped, have
been calculated.

\bibitem{LVO2} V. G. Zubkov, G. V. Bazuev, G. P. Shveikin,
Sov. Phys. Solid State (English Transl.) {\bf 18}, 1165 (1976).



\end{thebibliography}
\end{document}